\documentclass{ifacconf}

\usepackage{natbib}        % required for bibliography
\usepackage{amsmath,amssymb,amsfonts,bbm}
\usepackage{mathtools}
\usepackage{booktabs}
\usepackage{dsfont}

\usepackage[dvipsnames]{xcolor}
\newtheorem{lemma}{Lemma}

\newtheorem{remark}{Remark}

\newcommand{\sqr}{$\hfill\square$}

\newcommand{\mc}[1]{\mathcal{#1}}
\newcommand{\bb}[1]{\mathbb{#1}}

\DeclareMathAlphabet{\mymathbb}{U}{BOONDOX-ds}{m}{n}

\begin{document}
\begin{frontmatter}

\title{Receding Horizon Games for Modeling Competitive Supply Chains\thanksref{footnoteinfo}} 

\thanks[footnoteinfo]{This work is supported by the Swiss National Science Foundation via NCCR Automation (Grant Number 180545) and the Natural Sciences and Engineering Research Council of Canada (Reference Number RGPIN-2023-03257).}

\author[First]{Sophie Hall} 
\author[First]{Laura Guerrini}  
\author[First]{Florian D\"{o}rfler}  
\author[Second]{Dominic Liao-McPherson}

\address[First]{Automatic Control Laboratory, ETH Zürich, 
   Switzerland, (e-mail: \{shall, laurague, dorfler\}@control.ee.ethz.ch).}
\address[Second]{University of British Columbia, Vancouver, Canada, (e-mail: dliaomcp@mech.ubc.ca)}

\begin{abstract}  
The vast majority of products we use daily are supplied to us through complex global supply chains that transform raw materials into finished goods and distribute them to end consumers. This paper proposes a modeling methodology for dynamic competitive supply chains based on game theory and model predictive control. We model each manufacturer in the supply chain as a rational utility maximizing agent that selects their actions by finding an open-loop generalized Nash equilibrium of a multi-stage game. To react to competitors and the state of the market, every agent re-plans their actions in a receding horizon manner based on estimates of market and supplier parameters thereby creating an approximate closed-loop equilibrium policy. We demonstrate through numerical simulations that this modeling approach is computationally tractable and generates economically interpretable behaviors in a variety of settings such as demand spikes, supply shocks, and information asymmetry.
\end{abstract}


\end{frontmatter}
%===============================================================================

\section{Introduction}
Supply chains are complex socio-technical systems that transform raw materials into finished products and deliver these products to consumers. In modern market economies, supply chains are interconnected networks of suppliers, manufacturers, retailers, wholesalers, and consumers, each selfishly seeking to maximize their utility. Modeling and control of supply chains is crucial for tackling a variety of real-world challenges such as disruptions to product availability caused by supply shocks and labor shortages (e.g., due to the pandemic or geopolitical conflicts). 
% 
% minimizing whole-chain carbon emissions, and designing effective regulatory regimes. 
Supply chains are challenging to model since they are highly dynamic, subject to a variety of constraints, and comprised of multiple interlinked competitive actors. 
As a result, supply chains present a unique and challenging setting for control in socio-technical systems.
% (transport link capacity, warehouse space, regulatory)

Control engineering principles have been successfully applied to modeling and control of supply chains in previous works. Typical examples include inventory control, modeling industrial dynamics, and policy design for avoiding the \textit{Bullwhip Effect}~\citep{vassian1955application,towill1982dynamic,towill1996industrial}, mainly using frequency domain tools. Recent literature capitalizes on advances in optimal control, predominantly model predictive control (MPC), to tackle problems in real-time inventory management and supply chain optimization~\citep{braun2003model,perea2003model,subramanian2014economic, li2024mitigating}, remanufacturing
\citep{li2021stochastic}, and production and marketing decisions~\citep{pekelman1974simultaneous,levin2008risk}. 
A survey is given in~\cite{ivanov2018survey,sarimveis2008dynamic}.
% We direct the interested reader to the review papers~\citep{ivanov2018survey,sarimveis2008dynamic} for a more thorough survey.

The control-oriented works primarily focus on vertically integrated supply chains, where all agents are working toward a common goal, potentially in a decentralized way. However, in reality, there is competition at every stage of the chain starting from resource extraction, through production, wholesaling, distribution, and retail. Game theory is a powerful framework for modeling and analyzing supply chains with multiple agents that have conflicting interests~\citep{cachon2006game}. Network equilibrium models conceptualize supply chains as a network of interconnected producers, suppliers, and consumers~\citep{nagurney2002supply,nagurney2006supply,anderson2010price} and solve for (typically Nash) equilibria of network games by solving variational inequalities. These equilibria can computed for large networks using distributed algorithms~\citep{belgioioso2022distributed} and are flexible enough to model a wide variety of scenarios, e.g., they have been used to study the impact of waste from electric and electronic equipment disposal regulations~\citep{hammond2007closed} and the consequences of labor constraints caused by the Covid-19 pandemic~\citep{nagurney2021supply}. 

Dynamic game theory~\citep{bacsar1998dynamic} has been used to study a variety of scenarios involving dynamic multi-agent competition including optimal pricing and inventory replenishment policies~\citep{netessine2006inventory,kirman1974dynamic,bernstein2004dynamic}, the impacts of technology acquisition~\citep{gaimon1989dynamic}, and capacity investment decisions based on reputational effects~\citep{taylor2007supply}. Dynamic games involve multi-period decision-making and typically require solving for feedback Nash equilibria (Nash equilibria over policies) which are challenging to compute, limiting these methods to simple settings to remain analytically tractable. 

%\dlm{Still doesn't quite ``introduce'' RHGs, I think the issue is you put RHG material in this paragraph before you introduce it formally in the next one which made it awkward, I merged your text into the next paragraph to make it smoother}

Instead in this paper, we propose a method for modeling dynamic supply chains using receding horizon games by combining multi-period network equilibrium models with model predictive control. Specifically, we compute open-loop equilibria and introduce feedback by re-planning at every time step to numerically construct approximate feedback-equilibrium policies. These policies then serve to form a closed-loop model of the supply chain. This approach introduces feedback while retaining computational tractability (even for complex models with many agents~\citep{hall2022receding, hall2024stability}) and allows us to consider complex scenarios involving dynamics and delays, constraints, information asymmetry\footnote{In general, each agent can have different and imperfect estimates of the parameters of the other agents and of the system dynamics.}, external forcing, and forecasts that cannot be captured by either network equilibrium or traditional dynamic game theory approaches. We illustrate the efficacy of our method by studying demand and supply shocks, and the impact of information asymmetry and demonstrate that we are able to replicate a variety of economically relevant behaviors. Specifically we show: (i) that dynamic pricing can be used to manage demand and inventory, (ii) how dynamic effects propagate through the supply chain, and lastly (iii) that information offers a competitive advantage.

% \begin{itemize}
%   \item Supply chains are important
%   \item Control theory is commonly used to study single supply chains
%   \item Economic and other single-agent MPC strategies have been applied to supply chain optimization
%   \item Supply chains are usually competitive
%   \item Network games are used to study static competitive supply chains but aren't dynamic
%   \item Need to use dynamic game theory. GTP/GT-MPC is a powerful tool used in other competitive environments
%   \item \sh{This is messy, lets us capture information asymmetry }
% \end{itemize}

% \dlm{Framing: This is a cool way to model competitive behaviors}
% \begin{itemize}
% \item \sh{We can use these EMPC controllers to model relevant the supply chain, cool behaviors: interactions}
% \item \sh{Real world: information asymmetry based on parameter estimates}
% \item \sh{If it converges, what too}
% \end{itemize}

\textit{Notation:} %We denote the normal cone operator of a closed convex set $\Omega \subseteq \reals^n$ by $\nc_{ \Omega}:\Omega \mto \reals^n$. 
%For two matrices, $(A,B)$ denotes vertical concatenation.
We denote by $\bb{Z}_N= \{0,\dots, N\!-\!1\}$ the sequence of the first $N$ non-negative integers. For an index set $\mc V = \{1, \ldots, m\} \subset \mathbb{Z}$ and a collection of matrices/vectors $\{A^v\}$ the vertical stacking operator is $(A^v)_{v\in \mc V} = (A^1, \ldots, A^m)$ and  $A^{-v} = (A^v)_{v \in \mc{V} \setminus \{v\}}$  
and $\text{blkdiag}\{A^v\}_{v\in\mc{A}}$ denotes the block diagonal matrix with
$A^1 , ..., A^m$ on the main diagonal. For the generalized (i.e., constrained) game
\begin{align} \label{eq:game}
\forall v \in \mc{V}:~\min_{u^v} \left\{ J^v(u^v,u^{-v}) ~|~(u^v,u^{-v}) \in \mc C \right\},
\end{align}
with cost $J^v$ and convex, compact constraint set $\mc{C}$ the vector $\bar u = (\bar u^{v})_{v\in \mc{V}}$ is a Generalized Nash Equilibrium (GNE) of the game~\eqref{eq:game} if it satisfies
\begin{align*}
J^v(\bar u^v,\bar u^{-v}) \leq  
\min_{u^v}\left\{
J^v(u^v,\bar u^{-v}) ~|~(u^v,\bar u^{-v})\in \mc C
\right\}, \; \forall v\in \mc{V}
\end{align*}
and a variational GNE (v-GNE) of \eqref{eq:game} if it satisfies $ F(u) + \mc{N}_{\mc C}(u) \ni 0$ where $u = (u^v)_{v\in \mc V}$, $F(u) = (\nabla_{u^v} J^v(u^v,u^{-v}))_{v\in \mc V}$ is the pseudo-gradient map and $\mc{N}_{\mc C}$ is the normal cone map of $\mc C$ \citep{dontchev2009implicit}.

\section{Problem Setting}
\begin{figure}[htbp]
  \centering
  \includegraphics[width=0.85\columnwidth]{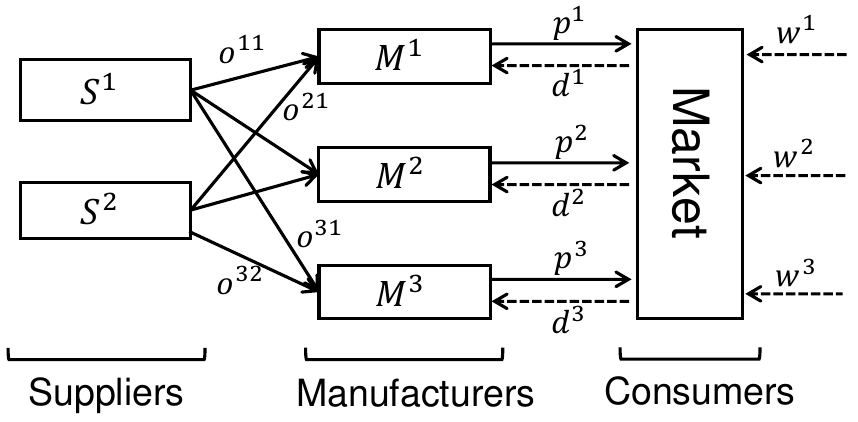}
  \caption{A supply chain consisting of two upstream suppliers, three manufacturers, and a competitive downstream market.}
  \label{fig:supply-chain}
\end{figure}

Consider a competitive supply chain consisting of $n_s > 0$ suppliers $\mc{S} = \{1,\ldots,n_s\}$, and $n_m > 0$ manufacturers $\mc{M} = \{1,\ldots,n_m\}$ serving a common market, as illustrated in Figure~\ref{fig:supply-chain}. The manufacturers transform raw material purchased from the suppliers into a finished product and sell it to consumers in the market. In this paper, we consider a product based on a single raw material.

The demand $d_t^v$ on day $t$ for the product of manufacturer $v\in \mc{M}$ is assumed elastic and set by the market according to the linear demand model~\citep{choi1991price}
\begin{equation}\label{eq:linear-demand}
  d_t^v = w_t^v - \beta^{vv} p_t^v + \sum_{j\in \mc{M}\backslash\{v\}} \beta^{vj} p_t^j,
\end{equation}
where $\beta^v = (\beta^{vj})_{j\in \mc{M}}\geq 0 $ are the coefficients of the demand model, $p^v_t \geq 0$ is the price chosen by manufacturer $v \in \mc{M}$, and $w_t^v$ is the base (or inelastic) demand. We observe that the demand for manufacturer $v$'s product falls when they increase their price $p^v$ and rises when their competitors increase their prices $p^{-v}$.

% \begin{remark}
% The demand model~\eqref{eq:linear-demand} can be rearranged:
% \begin{equation*}
% d^v_t = w_t^v - \underbrace{\Big(\beta^{vv} - \sum_{\scriptscriptstyle j\in \mc{M}\backslash\{v\}} \beta^{vj} \Big) p_t^v}_{\text{marginal consumer}}    + \underbrace{\sum_{\scriptscriptstyle  j\in \mc{M}\backslash\{v\}} \beta^{vj} (p^j_t - p^v_t)}_{\text{switching consumer}}
% \end{equation*}
% for $\beta^{vj}\to 0$ we only have marginal consumers, i.e., the product is completely differentiated and not substitutable and for $(\beta^{vv} - \sum_{j\not=v} \beta^{vj})\to 0$ there are only switching consumers, thus the products are substitutable and there is perfect competition~\citep{anderson2010price}. \sqr
% \end{remark}

To meet consumer demand, manufacturer $v\in \mc{M}$ orders $o_t^{vs}$ units of raw material from supplier $s\in \mc{S}$ and produces
\begin{equation} \label{eq:production}
  O_t^v = \sum_{s\in \mc{S}} o_t^{vs}
\end{equation}
units of product. They maintain an inventory $\xi_t^v$ of finished product in a warehouse and sell $d^v_t$ units of product to the common market at a price $p^v_t$. Each agent maintains enough inventory on hand to satisfy the demand (this will be enforced as a constraint later) and that both production and shipping are subject to a one day delay. This leads to the inventory balance equation
\begin{equation} \label{eq:inventory-balance}
  \xi_{t+1}^t = \alpha^v \xi_t^v + O_{t-1}^v - d^v_{t-1},
\end{equation}
where $\alpha^v \in (0,1]$ is the decay rate of the product ($\alpha^v = 1$ for non-perishable products)~\citep{schwager2016supply} and inventory is subject to the constraints $0\leq \xi_t^v \leq \Xi^v$ with maximum inventory $\Xi^v>0$. Inventory models of this type are common in supply chain analysis and control, e.g.,~\citep{braun2003model,spiegler2012control}. 

Each supplier $s\in \mc{S}$ can supply a maximum of $\bar O^s$ units per day, leading to the constraint
\begin{equation} \label{eq:supply-limit}
  \sum_{v \in \mc{M}} o_t^{vs} \leq \bar O^s,
\end{equation}
and sells raw material to manufacturer $v$ at the price 
\begin{equation}
  \rho^{s}(o_t^{vs}, o_t^{-vs}) = \rho_0^s + \rho_1^s \sum_{v\in \mc{M}} o_t^{vs}.
\end{equation}
This model of the wholesale price can be understood as an inverse form of the commonly used linear supply-price curve~\citep{schwager2016supply}. 
The first term is a constant minimum price set by each supplier where $\rho_0^s\geq0$. The second term introduces a dependency on the aggregate number of orders that can model a variety of commercial strategies from the suppliers. For example, setting $\rho_1^s > 0$ indicates the supplier feels they are in a position to raise prices when demand is high while conversely, setting $\rho_1^s < 0$ indicates the supplier is willing to share reductions in unit costs due to economies of scale.

By inserting~\eqref{eq:linear-demand} and~\eqref{eq:production} into~\eqref{eq:inventory-balance} and augmenting the state with the delay states, yields the following LTI dynamics of every manufacturer $v\in\mc{M}$:
\begin{subequations} \label{eq:agent-LTI}
\begin{equation} 
  x^v_{t+1} = A^v x^v_t + \sum_{j \in \mc{M}} B^{vj} u^j_t + D^v w^v_t,
\end{equation}
where $x^v_t = (\xi_t^v, O_{t-1}^v, d_{t-1}^v)$ represents the state, and $u^v_t = (o_t^{v1},\ldots, o_t^{vn_s},p^v_t)$ represents the control input of manufacturer $v$, while the base demand $w^v_t$ is an exogenous input, acting as a disturbance. The system matrices are:
\begin{gather}
  A^v = \begin{bmatrix}
    \alpha^v & 1 & -1\\
    0 & 0 & 0\\
    0 & 0 & 0
  \end{bmatrix},\;B^{vv} = \begin{bmatrix}
     0 & \cdots  & 0 & 0\\
    1 & \cdots & 1 & 0\\
    0 & \cdots & 0 & -\beta^{vv}
  \end{bmatrix},\;D^v = \begin{bmatrix}
    0 \\ 0 \\ 1
  \end{bmatrix},\nonumber\\
 \text{ and }  B^{vj} = \begin{bmatrix}
     0 & \cdots  & 0 & 0\\
    0 & \cdots & 0 & 0\\
    0 & \cdots & 0 & \beta^{vj}
  \end{bmatrix}\text{ for } j\in \mc{M}\backslash\{v\}.
\end{gather}
\end{subequations}
\begin{figure}
  \centering
  \includegraphics[width=0.9\columnwidth]{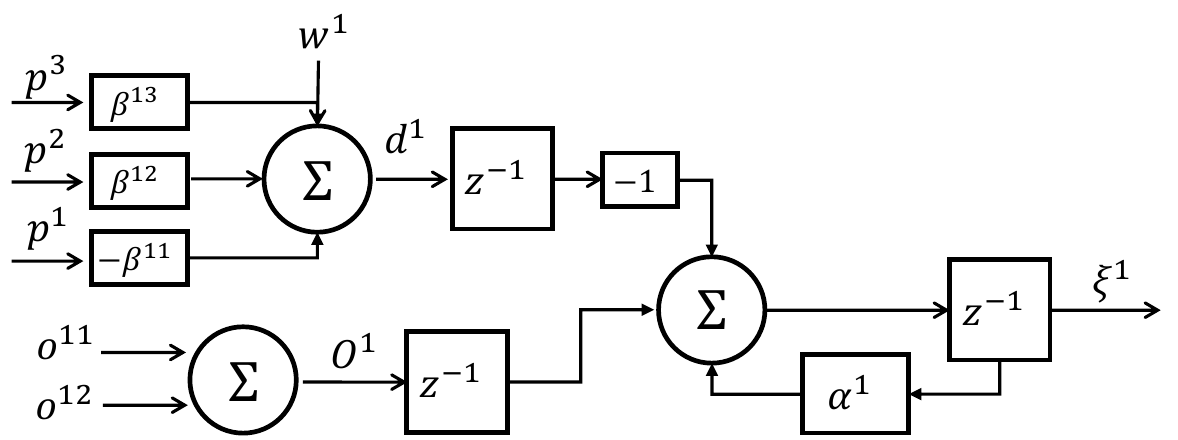}
  \caption{A block diagram of the inventory dynamics of $M^1$, where $z$ denotes the forward shift operator.}
  \label{fig:block-diagram}
\end{figure}
The dynamic model of the inventory balance~\eqref{eq:agent-LTI} for a single manufacturer $v$ is summarized in Figure~\ref{fig:block-diagram}. Collecting the inventory dynamics~\eqref{eq:agent-LTI}, we see that the global supply chain is governed by the following LTI system:
\begin{subequations} \label{eq:global-LTI}
\begin{equation} 
  x_{t+1} = A x_t + B u_t + D w_t
\end{equation}
with $ x_t = (x^v_t)_{v\in \mc{M}}$, $u_t = (u^v_t)_{v\in \mc{M}}$, $w_t = (w_t^v)_{v\in \mc{M}}$, and
\begin{equation} \label{eq:global-AB}
  A = \text{blkdiag}\{A^v\}_{v\in\mc{M}},~~ B = \begin{bmatrix}
    B^{11} & \cdots & B^{1n_m}\\
    \vdots & \ddots& \vdots\\
    B^{n_m 1} & \cdots & B^{n_m n_m}
  \end{bmatrix},
\end{equation}
\end{subequations}
and $D = \text{blkdiag}\{D^v\}_{v\in\mc{M}}$.

\begin{remark}
In this paper, we focus on a simple supply chain to illustrate the methodology. Single-agent MPC has been applied to complex multi-product/multi-layer networks, see, e.g.,~\cite{braun2003model,subramanian2014economic}. Our approach can be generalized similarly to add multiple intermediate suppliers and wholesalers etc. \sqr
\end{remark}

\section{Game-theoretic MPC-based modeling}
We want to model competitive dynamic supply chains. Therefore, we model each manufacturer as a rational agent seeking to minimize the following stage cost function:
\begin{multline}
  \ell^v(x^v,u^v,u^{-v},w^v) = \underbrace{\gamma^v \cdot (\xi^v - \bar \xi^v)^2}_{\text{security stock}} \\-\underbrace{\left (p^v\cdot d^v(p^v,p^{-v},w^v) - \sum_{s\in \mc{S}} \rho^s(o^{vs},o^{-vs})\cdot o^{vs}\right)}_{\text{net cash flow}}.
\end{multline}
The first term penalizes deviations from the safety inventory $\bar \xi^v$, which is a reference term used in inventory control laws (e.g.,~\citep{spiegler2012control}) to ensure that it is always possible to fulfill demand where $\gamma^v>0$, and the second term is net cash flow of manufacturer $v$.

%  \sh{I think is reasoning doesn't make complete sense, because it also penalizes if you have a lot of products and not only when having little} \dlm{agreed, some kind of linear penalty with cost of storage above and something else below would make more sense, just trickier to implement. E.g., 
% \begin{equation}
%   \begin{cases}
%   \gamma_1 (x - \bar x)  & x\geq \bar x\\
%   \gamma_2 (\bar x - x)^p & \text{else}
%   \end{cases}
% \end{equation}
% for $p = 1$ or $p = 2$ could be implemented using constraints and slacks, just not convinced its worth it}\sh{Agreed, not worth it for now!}

The manufacturers incorporate the dynamics of the supply chain into their decision-making process by minimizing their incremental cost function over a planning horizon $N > 0$. This yields the set of coupled optimal control problems (OCP) $\mc{P}_N(\theta,\mathbf x,w)$ for all $v\in \mc{M:}$ 
\begin{subequations} \label{eq:rhg-full}
\begin{alignat}{3}
&\min_{x^v,u^v}&&\sum_{k = 0}^{N-1} \ell^v(x_k^v, u_k^v,u_k^{-v},w_k^v) \label{eq:ocp-cost}\\
&\mathrm{s.t.} 
&&x^{v}_{k+1} = A^v x^v_k + \sum_{j\in \mc{M}} B^{vj} u_k^j + D^v w^v_k, \,\label{eq:ocp-dynamics}  k \in \bb{Z}_{N}  \\
& && x_k^v \in \mc{X}^v, \; k \in \bb{Z}_{N+1} \label{eq:ocp-state}\\
& && u_k^v\in \mc{U}^v, (u^v_k,u^{-v}_k) \in \mc{C} \label{eq:ocp-input},\; k \in \bb{Z}_{N} \\
& &&  x_0^v = \mathbf{x}^v\label{eq:ocp-initial},
\end{alignat}
where $\mathbf{x}^v$ is the initial condition, $u^v = (u^v_k)_{k = 0}^{N-1}$, and $x^v = (x^v_k)_{k = 0}^{N}$. The local constraint sets are $\mc{U}^v = \{(p^v, o^{vs})~|~0\leq p^v ,~0\leq o^{vs}~~\forall s\in \mc{S}\}$, $\mc{X}^v = \{(\xi,O,d)~|~0 \leq \xi \leq \Xi^v\}$
and
\begin{equation}
  \mc{C} = \left\{o^{vs}~ \bigg |~\sum_{v\in \mc{M}} o^{vs} \leq \bar O^s~~\forall s\in \mc{S} \right\}
\end{equation}
\end{subequations}
encodes the coupling constraint~\eqref{eq:supply-limit}. The parameters of the market, suppliers, and manufacturers are collected in $\theta = (\rho^s, \bar O^s, \beta^{v},\alpha^v, \gamma^v, \Xi^v)_{v\in \mc{M}, s\in \mc{S}}$. The demand forecast of each agent over the prediction horizon is $w^v = (w^v_k)_{k = 0}^N$. Together, the $n_p$ coupled OCPs $\mc{P}_N$ constitute a game.

In reality, there is information asymmetry (i.e., not all manufacturers have accurate information on the local parameters of others). To model these effects, we assume that each manufacturer is strategic, in the sense that they maintain models of their competitors. Specifically:
\begin{itemize}
  \item The form of the model is fixed and known; all agents plan over an $N$-step horizon;
  \item Each manufacturer maintains estimates of the game parameters $\hat \theta^v$ and the demand forecasts $\hat w^v = (\hat w^{vj})_{j\in \mc{M}}$ where $\hat w^{vj} = (\hat w^{vj}_k)_{k = 0}^N$ denotes the forecast of agent $v$ for agent $j$ over the prediction horizon $N$; 
  \item The state $x^v_t$ can be measured or estimated $\forall v\in \mc{V}$.% (or can be estimated).
\end{itemize}
The fact that each agent maintains a separate estimate of $\hat \theta^v$ and $\hat w^v$ can lead to information asymmetry, e.g., if agent 1 has a more accurate demand forecast than agent 2 they can exploit this asymmetry to obtain a competitive advantage (to be illustrated in Section~\ref{sec:demand-forecast}).

To compensate for the finite planning horizon and inaccurate information about each other, agents \textit{re-plan} their actions each day. Specifically, they find a Nash equilibrium $(\bar u^v_{k},\bar u^{-v}_{k})_{k = 0}^{N}$ of $\mc{P}_N(\hat \theta^v,x_t,\hat w^v_t)$ based on the measured state $x_t$, their parameter estimates $\hat \theta^v$ and their forecast $\hat w^v_t$ of the base demand over the interval $[t~~t+N]$, and apply the first element of their planned control sequence $u^v_t = \bar u^v_{0}(\hat \theta^v, x_t,\hat w^v_t)$. For each agent, this re-planning procedure constitutes a model predictive controller, wherein they use game theory to model how competitors will react to their actions (e.g., if I play $u^v_0$, rational competitors would play $u^{-v}_0$). Similar ideas are used in multi-robot systems, e.g., drone racing and autonomous driving~\citep{wang2021game,spica2020real,fridovich2020efficient} and energy management applications~\citep{hall2022receding}.

%We write~\eqref{eq:rhg-full} in a compact form by incorporating the dynamics~\eqref{eq:ocp-dynamics} into the cost function~\eqref{eq:ocp-cost}, and by substituting the constraints~\eqref{eq:ocp-state} into~\eqref{eq:ocp-input}."

We write~\eqref{eq:rhg-full} in a compact form by substituting the dynamics~\eqref{eq:ocp-dynamics} and initial condition~\eqref{eq:ocp-initial} into the cost~\eqref{eq:ocp-cost} and the constraints~\eqref{eq:ocp-state} resulting in a condensed game:
\begin{align} \label{eq:condensed-game}
\forall v\in \mc{V}:\quad \left\{
\begin{array}{rl}
 \displaystyle \min_{u^v}& {J^v(u^v,u^{-v}, \theta, \mathbf x, w^v)}\\
\textrm{s.t.} & G (\theta) u \leq g(\theta, \mathbf x,w),
\end{array}
\right. 
\end{align}
%
% \begin{subequations} \label{eq:condensed-game}
% \begin{alignat}{2}
% &\min_{u^v}~~&&J^v(u^v,u^{-v}, \theta, \mathbf x, w^v) \\
% &~\mathrm{s.t.} &&G (\theta) u \leq g(\theta, \mathbf x,w) \label{eq:condensed-cstr}
% \end{alignat}.
% \end{subequations}
where $u = (u^v)_{v\in \mc M}$, $\mathbf x = (\mathbf x^v)_{v\in \mc M}$, and $w = (w^v)_{k\in \mc M}$. The explicit definitions of $J^v, G$, $g$ are given in Appendix~\ref{sec:AppendixMatrices}. The feedback law $\kappa^v$ is given by
\begin{equation} \label{eq:rhg-policy}
  u^v_t = \kappa^v(x_t,\hat \theta^v, \hat w^v_t) = \Psi^v S(x_t, \hat \theta^v,\hat w^v_t)
\end{equation}
where $S(\mathbf x, \theta,w) = \{u~|~ u \text{ is a v-GNE of }~\eqref{eq:condensed-game}\}$ is the solution map of~\eqref{eq:condensed-game} and $\Psi^v$ is a matrix selecting $u^v_0$.% from $u$. 

Our global supply chain model is given by the supply chain dynamics~\eqref{eq:global-LTI} in feedback with the game-theoretic MPC policies~\eqref{eq:rhg-policy} yielding the following closed-loop dynamics
\begin{equation} \label{eq:closed-loop}
  x_{t+1} = A x_t + B \kappa(x_t, \hat \theta , \hat w_t) + D w_t
\end{equation}
where $\kappa(x_t, \hat \theta , \hat w_t) = (\kappa^v(x_t,\hat \theta^v,\hat w_t^v))_{v\in\mc M}$, and the collected parameter estimates and demand forecasts are denoted by $\hat w_t = (\hat w_t^v)_{v\in \mc M}$ and $\hat \theta = (\hat \theta^v)_{v\in \mc M}$ respectively.

We make several modeling decisions by defining our policies in this way:
(i) Choosing Nash over Stackelberg equilibria implies that all manufacturers operate at the same hierarchical level; (ii) We define closed-loop policies based on repeated open-loop equilibria instead of closed-loop feedback equilibria to ensure computational tractability. Despite recent advances (see e.g.,~\cite{laine2023computation, fridovich2020efficient}), only approximate methods exist to date for computing feedback equilibria. We demonstrate in Section~\ref{sec:numerical-studies} that RHG policies based on open-loop equilibria capture a variety of competitive behaviours. (iii) We do not consider collusion between manufacturers, this can be addressed in future works using coalitional game theory.
(iv) solving for \mbox{v-GNEs} ensures that the cost of enforcing the coupling constraints~\eqref{eq:ocp-input} is ``fairly'' distributed~\citep{belgioioso2022distributed}. 

%Our approach is algorithmic in the sense that we evaluate $\kappa^v$ by solving for v-GNEs of~\eqref{eq:condensed-game} numerically\footnote{Each agent maintains an estimate of all parameters and solves an independent instance of~\eqref{eq:rhg-full}, i.e., agents do not share information}. 
Each agent solves an independent instance of~\eqref{eq:rhg-full} based on their local estimates of parameters $\phi$ and demand $\hat{w}$ (i.e., information is not shared). Since the cost $J^v$ is quadratic and convex in $u^v$ for each agent if $\rho_1^s \geq 0$, v-GNEs of the condensed game~\eqref{eq:condensed-game} are computed by solving an affine variational inequality (AVI)~\citep{facchinei2010generalized}. The following Lemma, derived from~\citep[Theorem 2E.6]{dontchev2009implicit}, demonstrates that solutions to these AVIs are locally unique under mild conditions.

\begin{lemma}
Let $(\bar u, \bar \lambda)$ satisfy~\eqref{eq:KKT} and $\mc{A} = \{i~|~G_i \bar u = g_i(\bar x)\}$ denote the active constraints at $\bar u$. Then if $y^T H y > 0$ for all $y \in \{u~|~ G_i u = 0,~i\in \mc{A}\}$ and rank $((G_i)_{i\in \mc{A}}) = \textrm{cardinality}(\mc A)$, where $
  H u + f  = (\nabla_{u^v} J^v(\bar u^v,\bar u^{-v}, \hat \theta^v, \mathbf x))_{v\in \mc{M}}$ is the pseudo-gradient of~\eqref{eq:condensed-game}, then $(\bar u,\bar \lambda)$ is a locally unique \mbox{v-GNE} of~\eqref{eq:condensed-game}.
\end{lemma}
% \begin{remark}
% The explicit definitions of $J^v, G$, $g$, and the AVI for~\eqref{eq:condensed-game}, along with conditions for uniqueness are included in Appendix~\ref{sec:AppendixMatrices}. \sqr
% \end{remark}

 % (iii) the v-GNEs ``fairly'' distribute the cost of enforcing the coupling constraints~\eqref{eq:condensed-cstr}~\citep{belgioioso2022distributed}. 

% move to appendix!

\section{Numerical studies} \label{sec:numerical-studies}

%\dlm{An interesting question that popped into my head for the demand spike and supply shock (in the unlikely event you have time): What happens to net income?, i.e., does revenue go up more than cost and are the manufacturers better off before or after the shocks?}

We investigate the behavior of the dynamic supply chain model in four different settings to demonstrate that our methodology can capture interesting and interpretable competitive behaviors. 
%
%In study one and two we simulate demand and supply shocks, in study three we investigate what happens when agents have different forecasts of the demand $\hat w^v$, and in the final study we examine what happens under information asymmetry of the parameters. 
%
%In this paper, we only investigate scenarios where all manufacturers know the parameters perfectly (so that $\hat \theta^v = \theta$) and only consider information asymmetry in the demand forecasts $\hat w^v$. This can be relaxed in the future, for example, by having the manufacturers estimate each other's parameters using game-theoretic inference~\citep{maddux2023data,le2021lucidgames}.
%
The parameter values used in the simulations are listed in Table~\ref{tab:parameters}. The prediction horizon is chosen as $N = 15$ and the safety inventory and inventory limit is set to $\bar{\xi}^v = 25$ and $ \Xi^v = 50, \;\forall v \in \mc{M}.$ The initial condition for all simulations is $\mathbf{x}^v = \begin{bmatrix}
    30 & 0 & 0 
\end{bmatrix}^\top, \; \forall v \in \mc{M}.$ We solve for the \mbox{v-GNEs} of~\eqref{eq:condensed-game}  using a Smoothed Fischer{\textendash}Burmeister Method~\citep{liao2018regularized}.

\begin{table}[ht]
\begin{tabular}{@{}lllll@{}}
\toprule
$\alpha^v$  & $\beta^v = [\beta^{v1},\beta^{v2},\beta^{v3} ]$ &$\gamma^v$  & $\rho_0^s$ &$\rho_1^s$ \\ \midrule
$\alpha^1 = 0.9$& $\beta^1 = [0.7, 0.3, 0.3]$ & $\gamma^1 = 0.1$ & $\rho_0^1 = 1 $  & $\rho_1^1 = 0.1 $\\ 
  $\alpha^2 = 0.7$ &$\beta^2 = [0.3, 0.8, 0.3]$& $\gamma^2 = 0.1$  &$\rho_0^2 = 1.5 $  & $\rho_1^2 = 0.15 $   \\
  $\alpha^3 = 0.5$& $\beta^3 = [0.3, 0.3, 0.6]$ & $\gamma^3 = 0.1$& &  \\\bottomrule
\end{tabular}
\caption{Parameter values for all case studies.}\label{tab:parameters}
\end{table}

\subsection{Demand spike}
\label{sec:demand-spike}
\begin{figure}
  \centering
  \includegraphics[width=1\columnwidth]{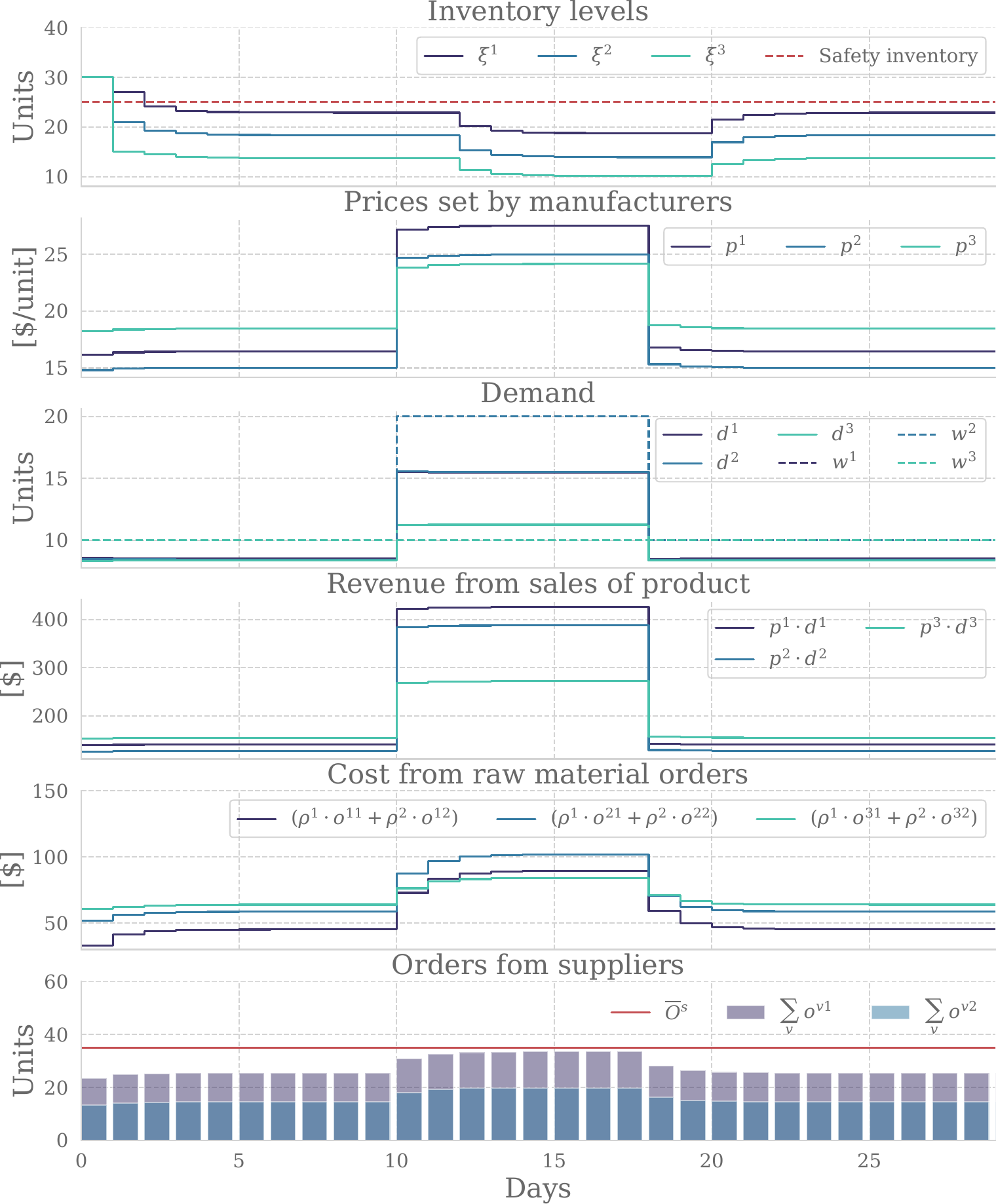}
  \caption{A spike in base demand $w_t^v$ by a factor of two for manufacturers $M^1$ and $M^2$ for $t\in \mathbb{Z}_{[10,18]}$. The manufacturers have no preview of the demand spike.}
  \label{fig:demand-spike}
\end{figure}
Consider a scenario where base demand suddenly spikes, as, e.g., during a pandemic. We simulate the spike by doubling the base demand $w_t^v, \forall\, v \in \{1,2\}$ for days $t\in \mathbb{Z}_{[10,18]}$ without giving the agents a preview of the spike. The results, shown in Figure~\ref{fig:demand-spike}, demonstrate that:

\begin{itemize}
  \item The manufacturers $M^1$ and $M^2$ exploit the demand spike to increase their prices and thus also increase their revenue (second, third, and fourth plot).
  \item The suppliers also get a share of ``the pie" as they can increase their wholesale price $\rho(o)$ with the increase in orders (bottom plot) and, as expected in a supply chain, the increased demand trickles-down.
  \item $M^3$ who does not have a demand spike also increases the prices to compensate for higher raw material cost. $M^3$ can increase prices without fearing loss of demand as others increased their prices. 
\end{itemize}
Overall, as expected, manufacturers not experiencing the demand spike also benefit economically, thanks to the expanded consumer base.

\subsection{Supply shock}
As a second case study, we investigate the global supply chain behavior in the presence of sudden shortages in supply of input material. This may occur if, e.g., supply routes are disrupted by geopolitical conflicts, labor shortages, or natural disasters such as droughts or flooding.

\begin{figure}
  \centering
  \includegraphics[width=1\columnwidth]{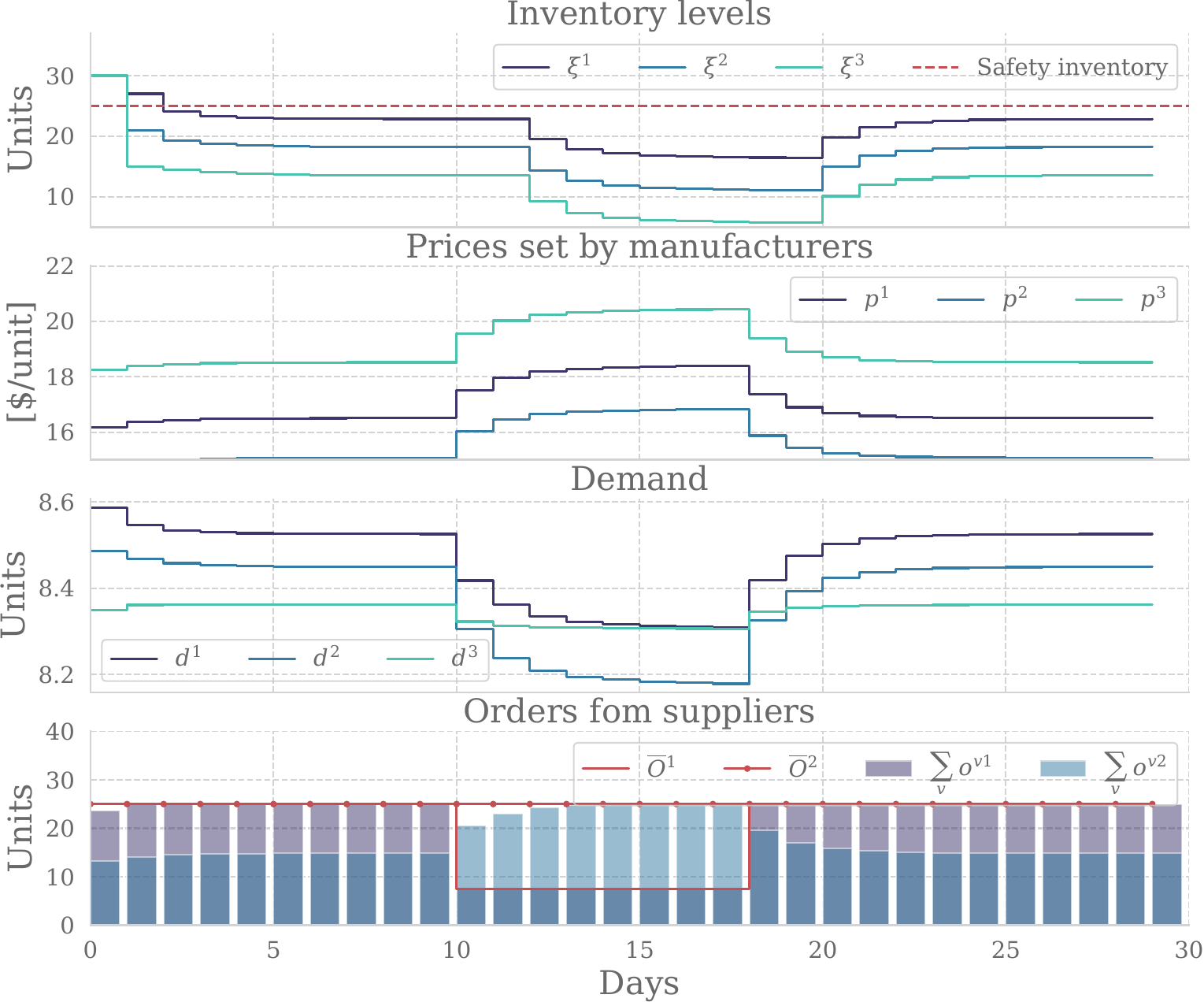}
  \caption{A sudden drop of $70\%$ in the supply capacity $\bar{O}^1$ of supplier $S^1$ for $t\in \mathbb{Z}_{[10,18]}$.}
  \label{fig:supply-shock}
\end{figure}

As in the previous case study, manufacturers are given no warning of the sudden 70\% drop in the supply limit for supplier one $\bar{O}^1$ on day 10. The results are displayed in Figure~\ref{fig:supply-shock}. We make the following key observations:

\begin{itemize}
    \item As manufacturers hit the supply limit $\bar{O}^1$ for the cheaper supplier $S^1$, they instantly order more from the other available supplier.
    \item To compensate for the increase in raw material cost, they increase their prices, experiencing a drop in demand as the number of marginal consumers decreases. 
    \item Manufacturers use prices to control demand (lower it) to avoid running out of inventory as they can not order enough raw material to keep up with demand.
\end{itemize}
\subsection{Information asymmetry: Demand forecast} \label{sec:demand-forecast}
In the real world, not all manufacturers have an accurate demand forecasts. To study information asymmetry with respect to $w_t^v$, we consider the same demand spike as in Subsection~\ref{sec:demand-spike} and investigate what happens when: %nor do they know what information is available to other manufacturers. 
\begin{enumerate}
    \item[(i)] $M^1$ has perfect forecast of $w_t$, and $M^2$ has no forecast;
    \item[(ii)] Both $M^1$ and $M^2$ have no forecast of $w_t$.
\end{enumerate}
The results are given in Figure~\ref{fig:inf-asymm} and we observe that:
\begin{figure}
  \centering
  \includegraphics[width=1\columnwidth]{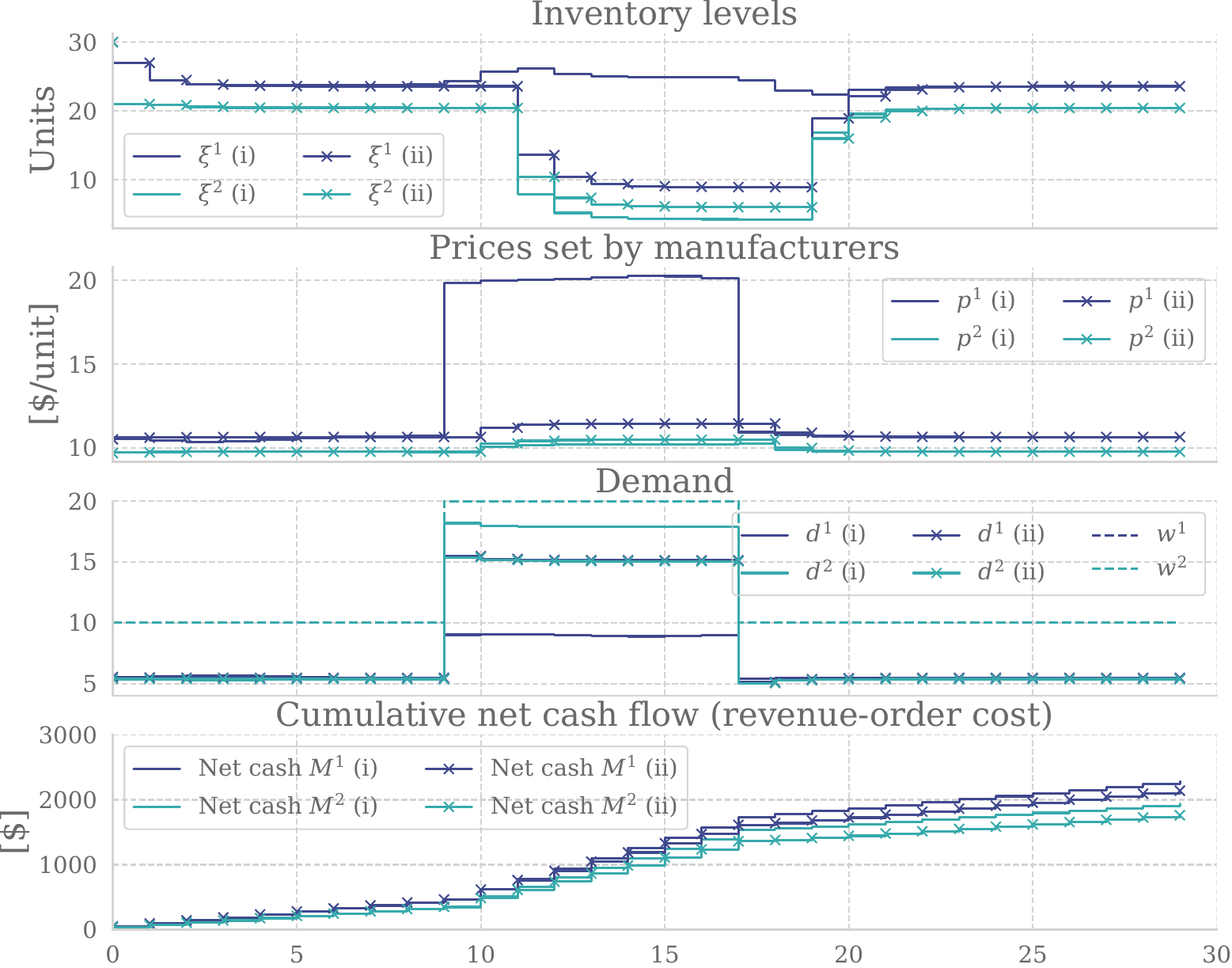}
  \caption{Information asymmetry: Case (i) $M^1$ has perfect forecast of the demand spike and $M^2$ has no forecast; Case (ii) none of the manufacturers have forecast.}
  \label{fig:inf-asymm}
\end{figure}
\begin{itemize}
    \item $M^1$ exploits the preview they have in case 1 in order to increase the price and therefore the net cash flow.
    \item $M^1$ uses the price inputs to control their inventory which, is cheaper than placing additional orders.
    \item $M^2$ suffers more when $M^1$ has the forecast due to the price-demand coupling between the manufacturers. 
\end{itemize}
Overall, from the net cash flow (bottom plot) we notice that $M^1$ gets a competitive advantage from having preview leading to better inventory control and higher profits.

\subsection{Information asymmetry: Market coupling}
\begin{figure}
  \centering
  \includegraphics[width=0.96\columnwidth]{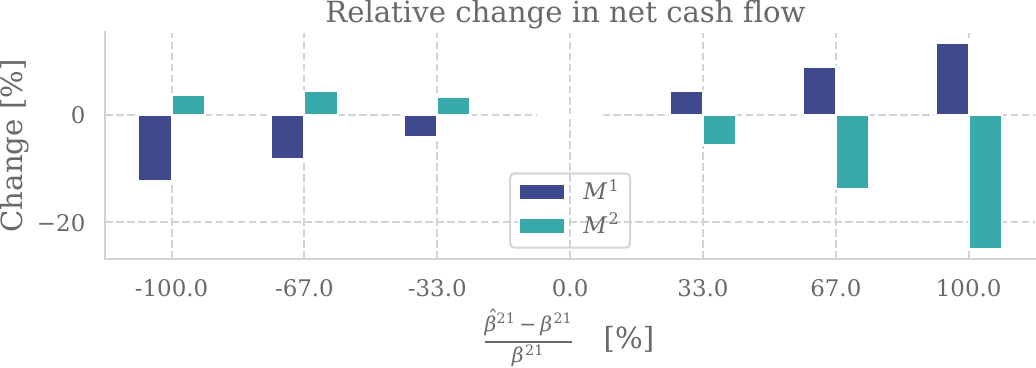}
  \caption{Economic impact of the parameter estimate $M^2$ has of $\hat{\beta}^{21}$  on both manufacturers.}
  \label{fig:inf-asymm-param1}
  \end{figure}

\begin{figure}
  \centering
  \includegraphics[width=0.96\columnwidth]{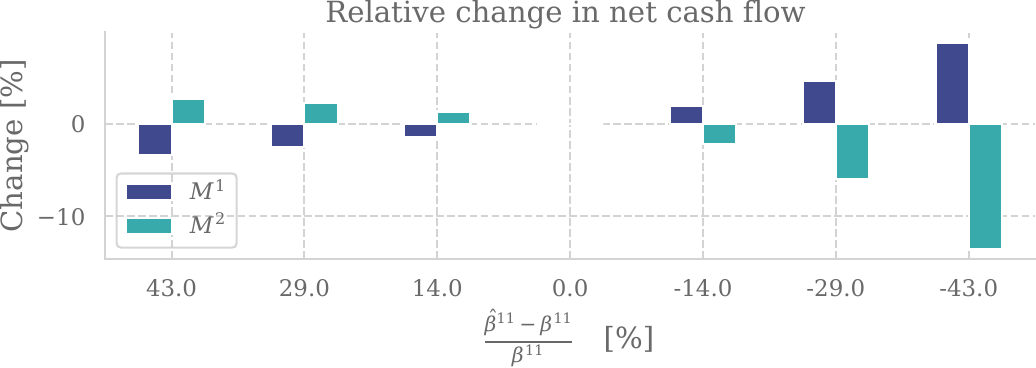}
  \caption{Change in net cash flow plotted against accuracy of parameter estimate $M^2$ has of $\hat{\beta}^{11}$.}
  \label{fig:inf-asymm-param2}
\end{figure}

We examine the economic impact of manufacturers misestimating the market coupling parameters $\beta^{vj}$, which influence the demand curve in~\eqref{eq:linear-demand}, particularly focusing on relative change in net cash flow in a two-manufacturer game. Specifically, we analyze scenarios where $M^1$ has perfect knowledge of all parameters and
\begin{enumerate}
    \item[(i)] $M^2$ either overestimates or underestimates $\beta^{21}$, assuming that the price set by $M^1$ has a greater or lesser impact on their own demand $d^2$ than it actually does,
    \item[(ii)] $M^2$ overestimates or underestimates $\beta^{11}$, misjudging the impact of $M^1$'s price $p^1$ on their demand $d^1$.
\end{enumerate}
The results are shown in Figure~\ref{fig:inf-asymm-param1} and \ref{fig:inf-asymm-param2}, we see that:
\begin{itemize}
    \item $M^2$ suffers economically by overestimating $\beta^{21}$ as they begin to act more conservatively. Conversely, $M^2$'s net cash flow changes only slightly when underestimating $\beta^{21}$, suggesting a benefit from assuming weaker coupling than present.
    \item $M^2$ suffers when overestimating $\beta^{11}$, thus $M^1$'s effect on their demand $d^1$, though less so than with the coupling parameter $\beta^{21}$, indicating that misestimating coupling terms has greater economic impact.
\end{itemize}

The study indicates that an agents internal model of their opponent has a significant impact on economic outcomes. In this scenario, it is better to underestimate the strength of the coupling terms and use a more aggressive strategy. However, this is scenario specific and better understanding the impact of parameter estimation on competitive performance is an important direction for future work.

% Studying the influence of approximate NEs under parameter inaccuracies on the outcome is necessary in future work.

% choice of parameter estimation significantly affects competition outcomes, suggesting economic benefits for agents considering weaker coupling. 


\textit{Summary}: In our four case studies, we observe the following economically relevant behaviors in our dynamic supply chain model: 1. Agents use prices to control demand and to manage their inventory; 2. increased demand and profits are passed throughout the supply chain; 3. information is a major competitive advantage;  4. agents without an acccurate demand forecast suffer from the information asymmetry; 5. Considering weaker coupling than actually present can lead to economically beneficial outcomes. 

 \subsection{Turnpike of the open-loop trajectories}
\begin{figure}[b]
  \centering
  \includegraphics[width=0.95\columnwidth]{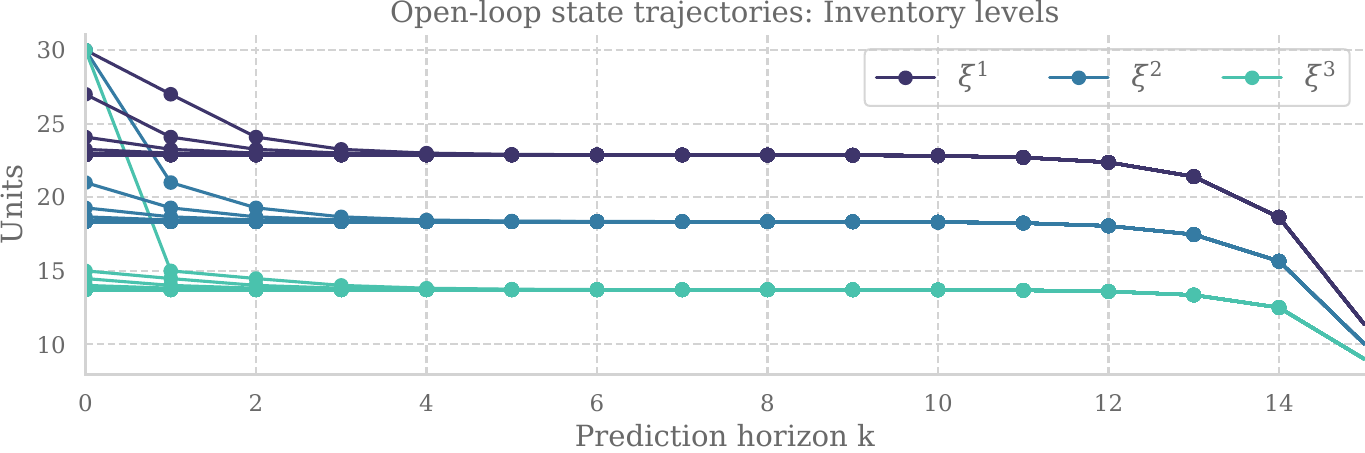}
  \caption{All open-loop trajectories over the simulation horizon of $t\in \mathbb{Z}_{30}$ days exhibit a turnpike effect.}
  \label{fig:turnpike}
\end{figure}

Besides the economically relevant behaviors, we also observe interesting system  properties RHG policies such as the turnpike property exhibited by the open-loop trajectories, see Figure~\ref{fig:turnpike}. The turnpike property is a well-known phenomenon in economic MPC where an optimal trajectory approaches an equilibrium state, stays close to it for a while, and then turns away from it again~\citep{damm2014exponential}. This observation calls for a detailed study of the system properties of RHG policies in future works.

\section{Conclusion}
We present a dynamic supply chain model in which manufacturers compete to maximize their profit by setting prices and managing their inventory levels. We formalize the problem as a game-theoretic MPC where the approximate equilibrium policies serve as a closed-loop model for the supply chain. We study the behavior of the supply chain in scenarios such as demand spikes, supply shocks, and demonstrate that economically intuitive behaviors emerge from our model. Future work includes: game-theoretic inference for validation and parameter estimation, studying recursive feasibility under information asymmetry, and using the model to study the design of market regulations and carbon reduction incentives. %understanding the impact of varying prediction horizons, as well as studying a variety of supply chain scenarios such as the design of market regulations and carbon reduction incentives.  

%and improving the algorithms used to solve the AVIs 

\section{Acknowledgements}
We we would like to thank Efe Balta and Giuseppe Belgioioso for their support in the early conception of the idea and and their support throughout the project. 
\appendix
\section{}\label{sec:AppendixMatrices} 
Variational GNE's of the condensed game~\eqref{eq:condensed-game} can be found by solving the following AVI (we have suppressed all dependencies on $\hat \theta^v$ for notational simplicity):
\begin{subequations} \label{eq:KKT}
\begin{align}
\begin{bmatrix}
  H & G^T\\
  -G & 0
\end{bmatrix} \begin{bmatrix}
  u \\ \lambda
\end{bmatrix} + \begin{bmatrix}
  f(\mathbf x)\\
  g(\mathbf x)
\end{bmatrix} + \begin{bmatrix}
\{0\}\\
\mc{N}_{\Lambda}(\lambda)  
\end{bmatrix} \ni 0
\end{align}
\end{subequations}
where $\mc{N}_\Lambda$ is the normal cone mapping of the set $\Lambda = \{\lambda ~|~ \lambda \geq 0\}$ and $ H u + f(\mathbf x)  = (\nabla_{u^v} J^v(u^v,u^{-v}, \hat \theta^v, \mathbf x))_{v\in \mc{M}}$ is the pseudo-gradient. The AVI~\eqref{eq:KKT} is of the same form as the KKT conditions for a quadratic program and can be readily solved using the algorithm in~\citep{liao2018regularized} with some small modifications to account for the asymmetry of $H$.

The objects $H = H_1 +  H_2$ and $f$ are defined as:
\begin{align*}
&H_1 = \left[\begin{smallmatrix}
    2R^{11} & \cdots & R^{1n_m}\\
    \vdots & \ddots& \vdots\\
    R^{n_m 1} & \cdots & 2R^{n_m n_m}
  \end{smallmatrix}\right], \\
  &R^{vv} = I_{N} \otimes  \left[\begin{smallmatrix}\text{blkdg}(\rho_1^s)& 0 & \\
0 & \beta^{vv} \end{smallmatrix}\right], R^{vj} = I_{N} \otimes  \left[\begin{smallmatrix}\text{blkdg}(\rho_1^s)& 0 & \\
0 & -\beta^{vj} \end{smallmatrix}\right],\quad
\end{align*}
and
\begin{align*}
&H_2=  \;\left[\begin{smallmatrix}
2(\tilde{B}^{11})^\top Q^1\tilde{B}^{11}&\hdots&(\tilde{B}^{11})^\top Q^1\tilde{B}^{1n_m}\\
 \vdots& \ddots&\vdots\\
 (\tilde{B}^{n_m1})^\top Q^{n_m}\tilde{B}^{n_m1}&\hdots& 2(\tilde{B}^{n_m1})^\top Q^{n_m}\tilde{B}^{n_m n_m}
\end{smallmatrix}\right],\\ 
&Q^v = I_{N} \otimes \left[\begin{smallmatrix}
        \gamma^v & 0 & 0\\
 0& 0 & 0 \\
 0 & 0 & 0 \\
    \end{smallmatrix}\right],
\end{align*}
and further
\begin{align*}
    f^v(\mathbf{x}^v) = (r^v +  (\tilde{B}^{vv})^\top Q^v (\tilde{A}^v\,\mathbf{x}^v + \tilde{D}^v\,w^v)   +  (\tilde{B}^{vv})^\top q^v)
\end{align*}
with $q^v = \mathds{1}_N \otimes \left[\begin{smallmatrix}(- 2\gamma^v\, \bar{\xi}^{v}) &0 & 0\end{smallmatrix}\right]^\top$, $r^v = (r^v_k)_{k = 0}^{N-1}$ and    \mbox{$r_k^v = \left[\begin{smallmatrix}
    \rho_{0,k}^\top & -w^v_k
\end{smallmatrix}\right]^\top$},  $\tilde{A}^v$,  $\tilde{B}^{vv}$,  $\tilde{B}^{vj}$ and $\tilde{D}^v$ are the $N$-step free-response, impulse-response and disturbance-response matrices respectively of the LTI system~\eqref{eq:agent-LTI} of agent~$v$. As in~\eqref{eq:linear-demand}, $w^v$ is the base demand of agent~$v$.
The constraint are encoded in $G$ and $g$ which can be constructed as follows:
\begin{align}
G =\left[\begin{smallmatrix}
        G_1\\
         G_2 \tilde{B}\\
          G_3\\
    \end{smallmatrix}\right], \quad g(\mathbf{x})= \left[\begin{smallmatrix}g_1\\ g_2- G_2(\tilde{A} \mathbf{x} + \tilde{D}\,w) \\ g_3\end{smallmatrix}\right],
\end{align}
with 
\begin{align*}
&G_1 = I_{N\cdot n_m}\otimes \left[\begin{smallmatrix}
-I_{n_s} & 0\\
 \mymathbb{0}_{n_s}& -1 \\
\end{smallmatrix}\right], \; g_1 = \mathds{1}_{N\cdot n_m} \otimes \left[\begin{smallmatrix} \mymathbb{0}_{n_s}\\  0 \\\end{smallmatrix}\right], \\
&G_2  =  I_{N\cdot n_m}\otimes  \left[\begin{smallmatrix}
  1 & 0 & 0 \\
   -1 & 0 & 0 \\
\end{smallmatrix}\right],\; g_2 = (g_2^v)_{v\in \mc{M}},\; g_2^v = \mathds{1}_N \otimes \left[\begin{smallmatrix} \Xi^v  \\ 0 \\\end{smallmatrix}\right],\\
&G_3 = \mymathbb{1}_{n_m}^\top\otimes (I_{N}\otimes \left[\begin{smallmatrix}
I_{n_s} & \mymathbb{0}_{n_s}\end{smallmatrix}\right]), \; g_3 = \mathds{1}_N\otimes [\bar{O}^1,\; \bar{O}^2]^\top,
\end{align*}
where $\tilde{A} = \text{blkdiag}\{\tilde{A}^v\}_{v\in\mc{M}}$, $\tilde{D} = \text{blkdiag}\{\tilde{D}^v\}_{v\in\mc{M}}$, and  $\tilde{B}$ defined as $B$ in \eqref{eq:global-AB} with $\tilde{B}^{jv}$ instead of ${B}^{jv}$. %
% \begin{align*}
%  \tilde{B} = \left[\begin{smallmatrix} 
%     \tilde{B}^{11} & \cdots & \tilde{B}^{1n_m}\\
%     \vdots & \ddots& \vdots\\
%     \tilde{B}^{n_m 1} & \cdots & \tilde{B}^{n_m n_m}
%  \end{smallmatrix}\right].    
% \end{align*}

% The v-GNEs 
%  have three important properties: i) They are unique under mild (and numerically verifiable) assumptions\footnote{Let $(\bar u, \bar \lambda) \in S(\mathbf x,w,\Theta)$ and $\mc{A} = \{i~|~C_i \bar u = g_i(\bar x)\}$ denote the active constraints at $\bar u$. Then if $y^T H y > 0$ for all $y \in \{u~|~ G_i u = 0,~i\in \mc{A}\}$ and rank $((C_i)_{i\in \mc{A}}) = |\mc A|$ then $(\bar u,\bar \lambda)$ is a locally unique v-GNE of~\eqref{eq:condensed-game}~\citep[Theorem 2E.6]{dontchev2009implicit}.}; ii) They can be computed efficiently by solving the affine VI~\eqref{eq:KKT} for which a variety of algorithms exits. Here we use an algorithm based on complimentary functions and a non-smooth Newton's method~\citep{liao2018regularized}, and iii) The v-GNEs ``fairly'' distribute the cost of enforcing the coupling constraints~\eqref{eq:condensed-cstr}~\citep{belgioioso2022distributed}.

\bibliography{RHGforSupplyChains}

\end{document}